\documentclass[a4paper,11pt]{article}
\usepackage{pos}

\title{Progress in top-quark pair production cross section calculations and impact on parton distribution functions of the proton}
\ShortTitle{Progress in top-quark pair production cross section calculations and impact on proton PDFs}

\author[a]{A. Ablat}
\author[a,b]{S. Dulat}
\author*[c]{M. Guzzi}
\author[d]{T.-J. Hou}
\author[c]{N. Kidonakis}
\author[a]{I. Sitiwaldi}
\author[e,c]{A. Tonero}
\author[b]{K. Xie}
\author[b]{C.-P. Yuan}

\affiliation[a]{School of Physics Science and Technology, Xinjiang University, Urumqi, Xinjiang 830046 China.}
\affiliation[b]{Department of Physics and Astronomy, Michigan State University, East Lansing, MI 48824, USA}
\affiliation[c]{Department of Physics, Kennesaw State University, Kennesaw, GA 30144, USA.}
\affiliation[d]{School of Nuclear Science and Technology, University of South China, Hengyang, Hunan 421001, China.}
\affiliation[e]{ Department of Chemistry and Physics, Florida Gulf Coast University, Fort Myers, FL 33965, USA}

\emailAdd{mguzzi@kennesaw.edu}
\emailAdd{alimablat@stu.xju.edu.cn}
\emailAdd{sdulat@hotmail.com}
\emailAdd{tjhou@msu.edu}
\emailAdd{nkidonak@kennesaw.edu}
\emailAdd{ibrahim010@sina.com}
\emailAdd{atonero@fgcu.edu}
\emailAdd{xiekepi1@msu.edu}
\emailAdd{yuanch@msu.edu}

\abstract{We discuss the impact of eligible top-quark pair production differential cross-section measurements at the LHC with a collision energy of 13 TeV on the parton distribution functions (PDFs) of the proton as well as the impact of approximate next-to-next-to-next-to-leading order (aN$^3$LO) QCD corrections combined with next-to-leading order (NLO) electroweak (EW) corrections on $t\bar t$ observables. 
We illustrate the effects on the gluon PDF at large $x$ from an optimal baseline selection of data in NNLO global fits, and show comparisons between the theory prediction for $t\bar t$ total and differential cross sections at aN$^3$LO QCD combined with NLO EW and recent measurements from the ATLAS and CMS collaborations at the LHC.}

\FullConference{12th Edition of the Large Hadron Collider Physics Conference (LHCP2024)\\
 3-7 June 2024\\
Boston, USA\\}

\renewcommand{\hookAfterAbstract}{%
\par\bigskip
\textsc{preprint}:
\href{MSUHEP-24-021}{MSUHEP-24-021}
}

\begin{document}
\maketitle

\section{Introduction}
The next precision frontier of Run-3 of the Large Hadron Collider (LHC) and beyond requires high-precision and high-accuracy theory predictions for particle reaction cross sections in perturbation theory to guarantee meaningful comparisons between theory and experiment.    
To this end, a combined effort in advancing the current knowledge of the structure of the proton encoded in the parton distribution functions (PDFs) as well as progress in higher-order computations in perturbation theory is indispensable.
The QCD factorization formula describing proton-proton collisions at the LHC depends on two fundamental ingredients: the PDFs of the proton, that are non-perturbative objects determined through global QCD analyses of world data, and the short-distance hard scattering contributions which are obtained using a variety of multi-loop techniques for Feynman diagram calculations in relativistic quantum field theories. 
Top-quark pair production cross section measurements are a clean probe for PDFs at intermediate and large longitudinal momentum fraction $x$ where they are currently poorly constrained and where jet-production measurements may complement those from $t \bar t$ in absence of tensions between these measurements. 
The impact of recent $t\bar t$ high-precision measurements at ATLAS and CMS on global QCD analyses of CTEQ PDFs at next-to-next-to-leading order (NNLO) is discussed in Ref.~\cite{Ablat:2023tiy}, while a recent study of the impact of corrections from soft-gluon contributions at approximate next-to-next-to-next-to-leading order (aN$^3$LO) in QCD combined with electroweak (EW) corrections at next-to-leading order (NLO) are discussed in Ref.~\cite{Kidonakis:2023juy}.

\section{Gluon impact from optimal combinations of 13 TeV top-quark measurements}
\begin{figure}[t]
\centering
\includegraphics[width=0.49\textwidth]{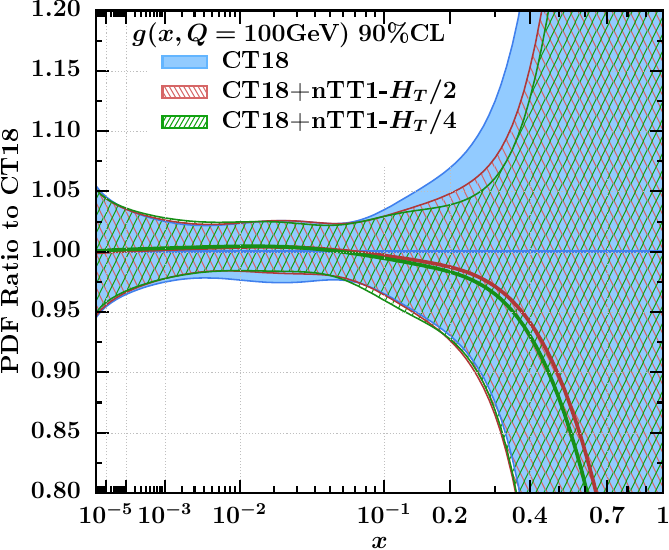}
\includegraphics[width=0.49\textwidth]{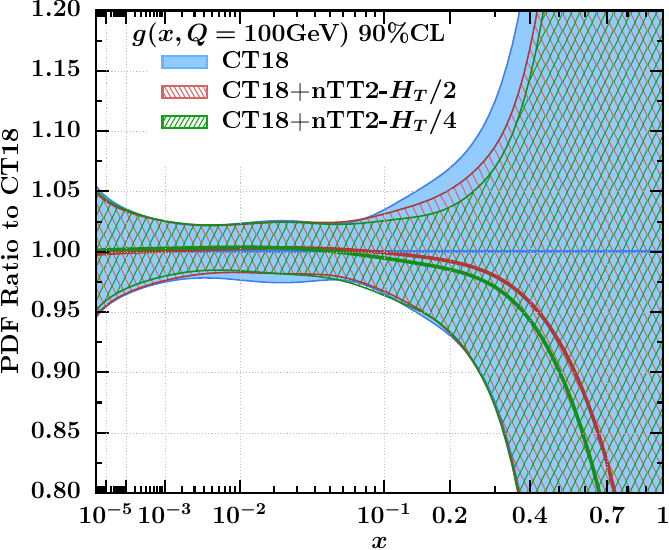}\\ (a) \hspace{3in}
(b)
\caption{Gluon PDFs and uncertainties. In (a) the CT18+nTT1 and (b) CT18+nTT2 global QCD analyses at NNLO, plotted as ratios to CT18 NNLO. PDF uncertainties are evaluated at the $90\%$ CL.}
\label{combin}
\end{figure}

In this section, we report the results published in Ref.~\cite{Ablat:2023tiy}. 
Future CTEQ PDF releases require substantial efforts in selecting sensitive data from the large amount of novel high-precision measurements at the LHC. In particular, we studied the impact of single differential cross section measurements of top-pair production at the LHC at 13 TeV from the ATLAS~\cite{ATLAS:2019hxz,ATLAS:2020ccu} and CMS~\cite{CMS:2018adi,CMS:2021vhb} collaborations on the gluon and other CTEQ PDFs~\cite{Ablat:2023tiy}. The PDF impact from these measurements has been analyzed by including each individual measurement on top of the CT18 baseline using the \texttt{ePump}~\cite{Schmidt:2018hvu,Hou:2019gfw} code first for a rapid assessment, and then in individual global fits. In such global fits, we analyzed the effect of including statistical correlations when these were available in the case of combined measurements (see Refs.~\cite{Ablat:2023tiy,ATLAS:2020ccu}), the impact from including the same distribution but with different binning, the interplay between top-quark and jet production data, the dependence on the factorization, renormalization and top-quark mass $m^{(pole)}_t$ scales, and EW radiative corrections at NLO when available.        
The NNLO QCD theory predictions are obtained through \texttt{fastNNLO} tables~\cite{Britzger:2015ksm,Czakon:2017dip,repo,Czakon:2016dgf} as well as NNLO/NLO $K$-factors obtained with the \texttt{MATRIX} code~\cite{Grazzini:2017mhc,Catani:2019iny,Matrixrepo} 
and NLO APPLgrid tables~\cite{Carli:2010rw} with \texttt{MCFM}~\cite{Campbell:2015qma}. EW corrections at NLO, either taken from Ref.~\cite{Czakon:2017dip} or calculated with \texttt{MadGraph\_aMC@NLO}~\cite{Pagani:2016caq}/\texttt{MCFM}~\cite{Campbell:2016dks} were also considered when available.
The most sensitive PDF information is obtained using two optimal combinations of measurements labelled CT18+nTT1 and CT18+nTT2 which minimize the tensions between data in the extended baseline, maximize the sensitivity and have optimal $\chi^2/N_{pt}$.       
CT18+nTT1 includes the rapidity distribution of the top-quark pair $y_{t\bar t}$ at ATLAS in the hadronic~\cite{ATLAS:2019hxz} and lepton+jet~\cite{ATLAS:2020ccu} channels and at CMS in the dilepton channel~\cite{CMS:2018adi}, and the invariant mass distribution of the top-quark pair $m_{t\bar t}$ at CMS in the lepton+jet channel~\cite{CMS:2021vhb}. CT18+nTT2 includes the same distributions except for the ATLAS lepton+jet $y_{t\bar
  t}$ that is replaced by the $y_{t\bar
  t}$ + $y^B_{t\bar t}$ + $m_{t\bar t}$ + $H^{t\bar t}_T$ combination. Here, $H_T^{t\bar{t}}$ is
the scalar sum of the transverse momenta of the hadronic and leptonic top quarks, and $y^B_{t\bar t}$ is the rapidity distribution for the boosted topology. 
The impact on the CTEQ gluon from the global fit with the CT18+nTT1 and CT18+nTT2 extended baselines is illustrated in Fig.~\ref{combin}. Hatched error bands represent the $H_T/2$ (red) and $H_T/4$ (green) choices for the central scale in the 13 TeV $t \bar t$ theory predictions. Most of the impact from the new data is visible in the large $x$ region and mostly due to the high-luminosity 137 fb$^{-1}$ data at CMS. The overall quality-of-fit of the CT18+nTT1 and CT18+nTT2
fits is essentially the same as that of CT18, with $\chi^2/N_{pt}\approx 1.16$. 
\section{Top-quark cross sections and distributions at aN$^3$LO}
In this section, we report the results published in Ref.~\cite{Kidonakis:2023juy}. 
Top-quark physics plays a central role in current and future research programs at the LHC. In fact, $t\bar t$ production is one of the most important processes at this collider; it has the ability to constrain proton PDFs and is important for searches of new physics interactions. 
Measurements of top-quark pair production total and differential cross section are going to be delivered with unprecedented precision and accuracy at the LHC in the near future. Therefore, it is critical to push forward the precision frontier of the theory calculations predicting this standard-candle observable. 
Theory predictions for top-quark cross sections and differential distributions in $t{\bar t}$ production have been computed at NLO ${\cal O}(\alpha_s^3)$ in Refs.~\cite{Nason:1987xz,Beenakker:1988bq,Nason:1989zy,Meng:1989rp,Beenakker:1990maa,Mangano:1991jk}, and NNLO at ${\cal O}(\alpha_s^4)$ in Refs.~\cite{Barnreuther:2012wtj,Czakon:2012zr,Czakon:2012pz,Czakon:2013goa,Czakon:2015owf,Czakon:2016ckf,Grazzini:2017mhc,Catani:2019iny,Catani:2019hip,Catani:2020tko} in QCD, while EW corrections to this process were studied in Refs.~\cite{Beenakker:1993yr,Bernreuther:2005is,Kuhn:2005it,Bernreuther:2006vg,Kuhn:2006vh,Hollik:2007sw,Bernreuther:2008md,Bernreuther:2010ny,Hollik:2011ps,Kuhn:2011ri,Manohar:2012rs,Bernreuther:2012sx,Kuhn:2013zoa,Campbell:2015vua,Campbell:2016dks,Pagani:2016caq,Denner:2016jyo,Czakon:2017wor,Carrazza:2020gss}.  
Logarithmic enhancements from soft-gluon emissions provide an important subset of QCD corrections which are dominant in the LHC kinematic regime and have been extensively studied in the literature in the past three decades~\cite{Sterman:1986aj,Catani:1989ne,Kidonakis:1996aq,Kidonakis:1997gm,Bonciani:1998vc,Laenen:1998qw,Kidonakis:2000ui,Bauer:2000yr,Kidonakis:2001nj,Bauer:2001yt,Beneke:2002ph,Kidonakis:2003qe,Kidonakis:2008mu,Kidonakis:2009ev,Czakon:2009zw,Ahrens:2009uz,Ahrens:2010zv,Aliev:2010zk,Kidonakis:2010dk,Ahrens:2011mw,Kidonakis:2011zn,Ahrens:2011px,Beneke:2011mq,Cacciari:2011hy,Czakon:2011xx,Ferroglia:2012ku,Beneke:2012wb,Ferroglia:2012uy,Kidonakis:2012rm,Ferroglia:2013awa,Kidonakis:2014isa,Kidonakis:2014pja,Kidonakis:2015ona,Kidonakis:2019yji,Kidonakis:2023juy}.  
Perturbative QCD corrections from soft gluons at third order in perturbative QCD were calculated in Refs.~\cite{Kidonakis:2014isa,Kidonakis:2014pja} based on
the formalism of Refs.~\cite{Kidonakis:1996aq,Kidonakis:1997gm,Kidonakis:2000ui,Kidonakis:2009ev,Kidonakis:2010dk,Kidonakis:2011zn,Kidonakis:2012rm,Kidonakis:2014isa,Kidonakis:2014pja,Kidonakis:2015ona,Kidonakis:2019yji}. In Ref.~\cite{Kidonakis:2023juy} these corrections are added on top of the exact NNLO cross section to obtain QCD results at approximate N$^3$LO (aN$^3$LO) which are combined with Electroweak (EW) corrections included at NLO.  
This calculation is used to obtain new results for the top-quark pair production total and differential cross sections which are compared to recent measurements from the LHC.  
The expression of the partonic contributions from soft-gluon corrections in QCD is in the form of $C_{ij}^{(k)} \left[\ln^k(s_4/m_t^2)/s_4\right]_{+}$.
In general, the coefficients $C^{(k)}_{ij}$, with $k \leq 5$ at N$^3$LO, depend on the Mandelstam variables $s$, $t_1$, $u_1$, the top-quark mass $m_t$, and the renormalization and factorization scales $\mu_R$ and $\mu_F$ respectively. The threshold variable $s_4$ is defined by $s_4=s+t_1+u_1$. More details about the these corrections can be found in Refs.~\cite{Kidonakis:2014isa,Kidonakis:2014pja}.

\begin{figure}
\begin{center}
\includegraphics[width=125mm]{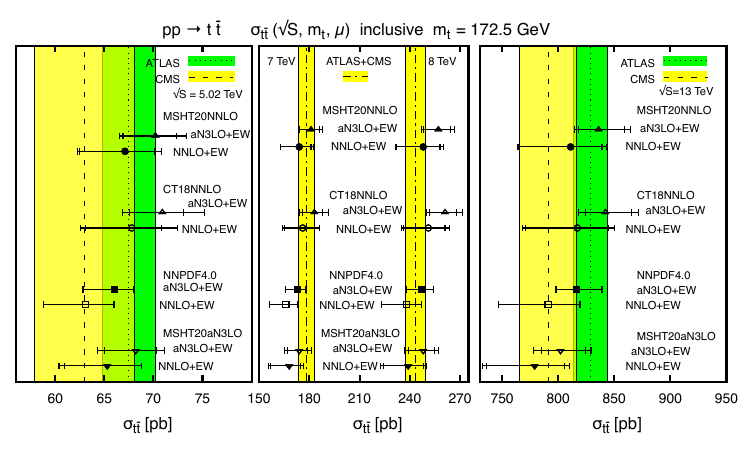}
\caption{$t \bar t$ total inclusive cross sections compared to recent measurements at the LHC at different collision energies.} \label{Xsec-Tot}
\end{center}
\end{figure}

\begin{figure}[htbp]
\begin{center}
\includegraphics[width=140mm]{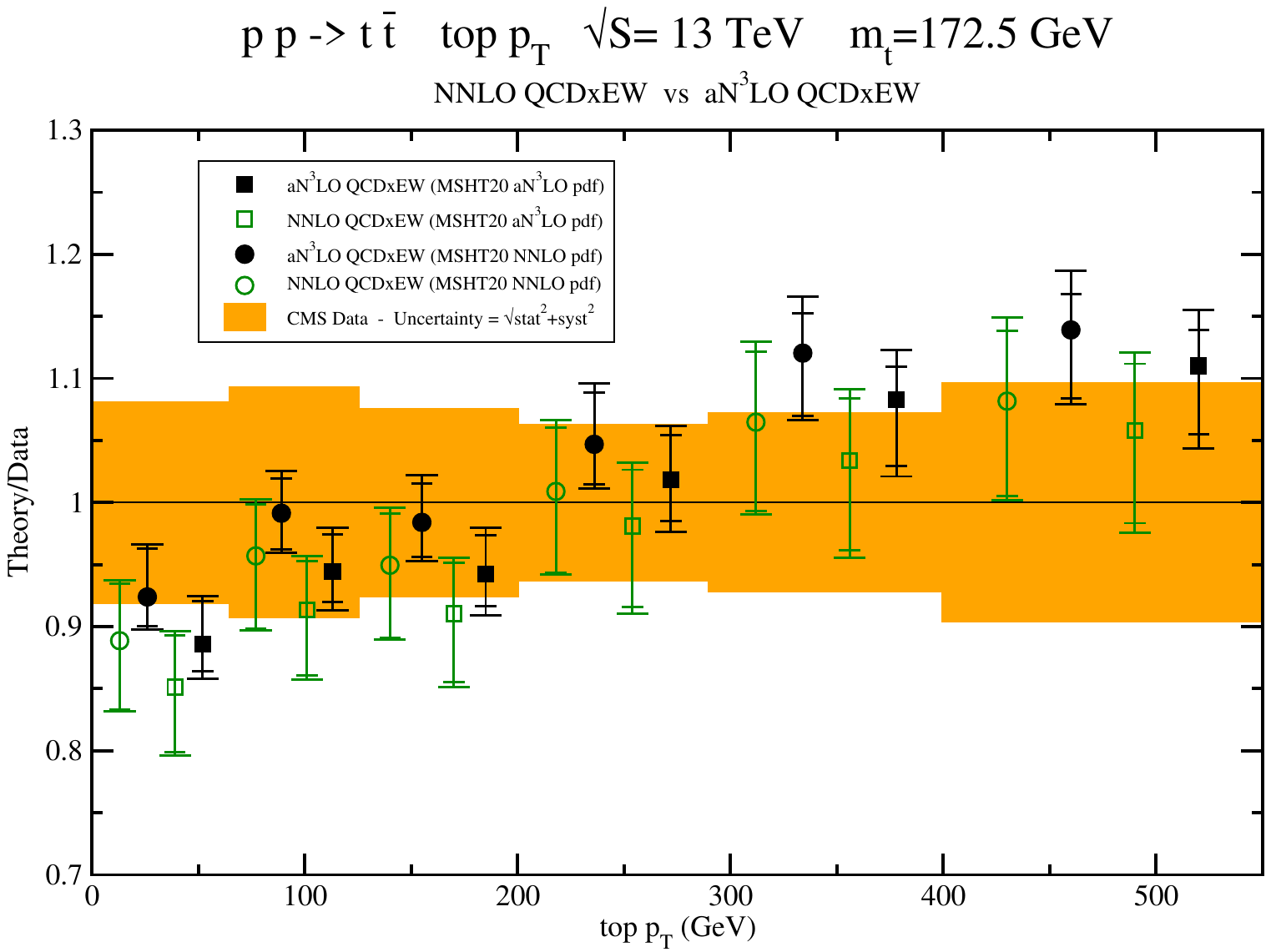}
\includegraphics[width=140mm]{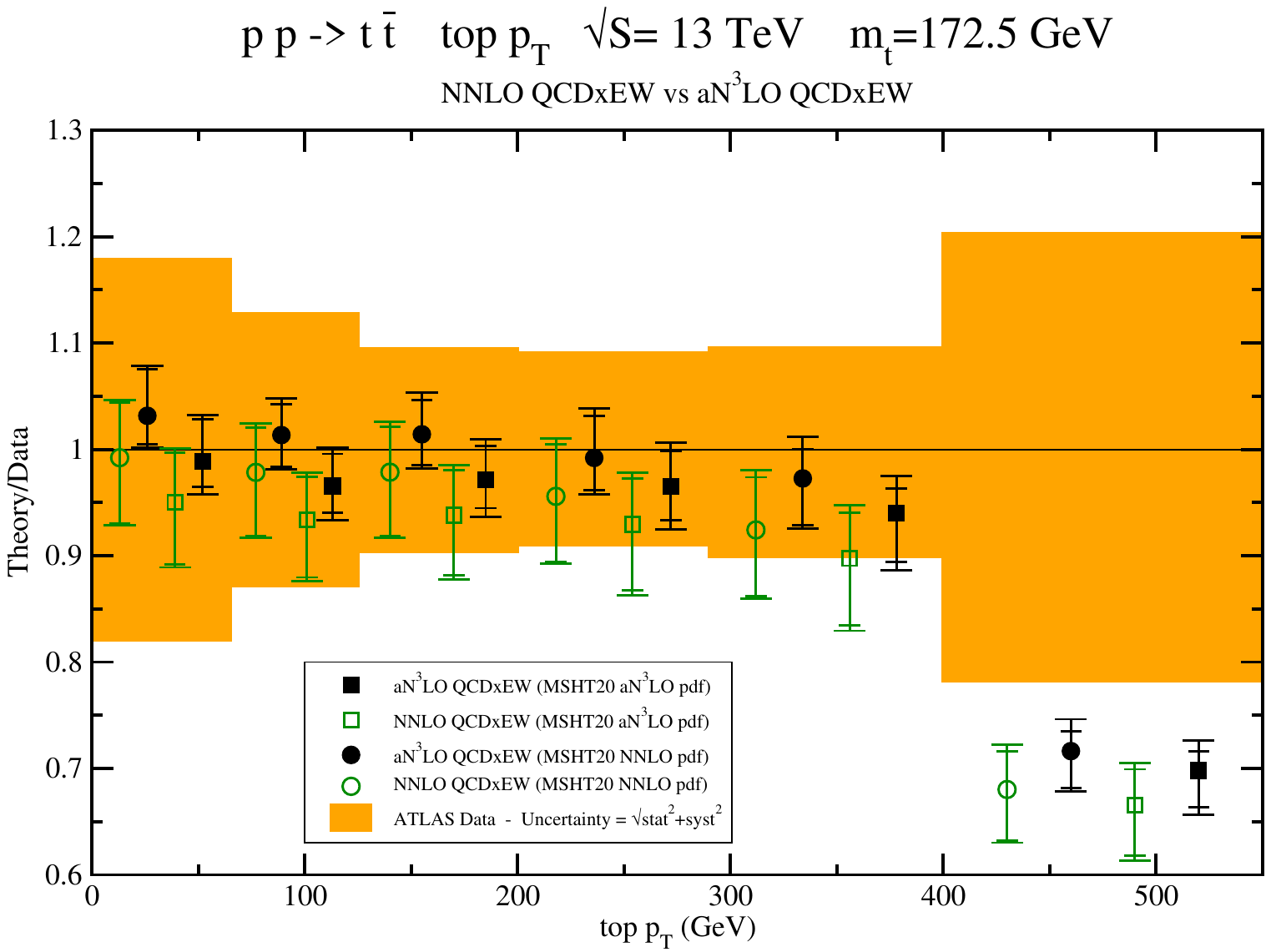}
\caption{Comparison of NNLO QCD$\times$EW and aN$^3$LO QCD$\times$EW theory predictions using MSHT20 NNLO and aN$^3$LO PDF with CMS (upper plot) and ATLAS (lower plot) top-quark transverse momentum  data.}
\label{pT-ditr}
\end{center}
\end{figure}

In Figure~\ref{Xsec-Tot}, we compare our theoretical predictions at aN$^3$LO obtained with different PDFs  
(CT18~\cite{Hou:2019efy}, MSHT20~\cite{Bailey:2020ooq,McGowan:2022nag}, and NNPDF4.0~\cite{NNPDF:2021njg}) 
to recent LHC measurements for various collision energies such as ATLAS~\cite{ATLAS:2022jbj} and CMS~\cite{CMS:2021gwv} at $\sqrt{S}=5.02$ TeV; ATLAS and CMS combined at $\sqrt{S}=7$ and 8 TeV~\cite{ATLAS:2022aof}; ATLAS~\cite{ATLAS:2023gsl} and CMS~\cite{CMS:2021vhb} at $\sqrt{S}=13$ TeV. 
The theory error bars represent scale uncertainty (inner bar), and scale added to PDF uncertainties in quadrature (outer bar). The experimental error bands represent all the given errors added in quadrature. The NNLO theory predictions are calculated using \texttt{Top++2.0}~\cite{Czakon:2011xx} while the NLO QCD+EW are obtained with \texttt{MadGraph5\_aMC@NLO}~\cite{Alwall:2014hca,Frederix:2018nkq}. More results are available in Ref.~\cite{Kidonakis:2023juy}. 
In Figure~\ref{pT-ditr}, we illustrate ratio plots where we compare the theory results for the top-quark transverse momentum $p^t_T$ distribution at NNLO and aN$^3$LO with EW corrections to recent measurements at ATLAS~\cite{ATLAS:2019hxz} and CMS~\cite{CMS:2018adi} at 13 TeV. The orange band represents the sum of statistical and systematic experimental uncertainties added in quadrature. Inner (outer) bars represent scale (scale plus MSHT20 PDFs at aN$^3$LO or NNLO, respectively) theoretical uncertainties. The NLO EW corrections are obtained from Ref.~\cite{Czakon:2017wor} while the combined QCD$\times$EW corrections incorporate ${\cal O}(\alpha_s^2\alpha)$ terms and the subleading ${\cal O}(\alpha_s\alpha^2)$, ${\cal O}(\alpha^3)$, ${\cal O}(\alpha_s^3\alpha)$ terms which are included using the multiplicative method discussed in Ref.~\cite{Czakon:2017wor}. Similar results with other PDFs as well as for the rapidity distributions of the top quark are presented in Ref.~\cite{Kidonakis:2023juy}.    

\section{Conclusions}
We reported recent progress in top-quark pair production cross section calculations and presented a comprehensive study of the impact of recent high-precision measurements from ATLAS and CMS at the LHC at 13 TeV on CTEQ PDFs. We identified two optimal combinations of $t\bar t$ measurements that are added on top of the CT18 baseline and  maximize the information extracted from the data. Both combinations have mild impact on the CT18 gluon and overall have the same fit quality of CT18. 

We studied the impact of radiative corrections at aN$^3$LO in QCD from soft-gluon resummation including EW corrections at NLO on total and differential $t\bar t$ cross sections. These new predictions have been compared to recent $t\bar t$ cross section measurements at the LHC, and the QCD corrections substantially increase the rates for $\sigma_{t\bar t}$ and the differential distributions while the EW corrections play an important role at large $p^t_T$. 

\acknowledgments
The work of A.A., S.D. and I.S. is supported by the National Natural Science Foundation of China under Grants No. 11965020 and No. 11847160. 
The work of M.G., N.K. and A.T. has been supported by the National Science Foundation under Grants No.~PHY 2112025 and No.~PHY 2412071. 
The work of T.-J. Hou was supported by Natural Science Foundation of Hunan province of China under Grant No. 2023JJ30496. 
C.-P.Y. and K.X. at MSU are supported by 
the U.S. National Science Foundation under Grants No.~PHY 2310291 and PHY 2310497.
This work used resources of high-performance computing clusters from MSU HPCC, KSU HPCs, as well as Pitt CRC.

\newpage
\bibliographystyle{JHEP}

\begin{thebibliography}{10}

\bibitem{Ablat:2023tiy}
A.~Ablat, M.~Guzzi, K.~Xie, S.~Dulat, T.-J. Hou, I.~Sitiwaldi et~al.,
  \emph{{Exploring the impact of high-precision top-quark pair production data
  on the structure of the proton at the LHC}},
  \href{https://doi.org/10.1103/PhysRevD.109.054027}{\emph{Phys. Rev. D}
  {\bfseries 109} (2024) 054027}
  [\href{https://arxiv.org/abs/2307.11153}{{\ttfamily 2307.11153}}].

\bibitem{Kidonakis:2023juy}
N.~Kidonakis, M.~Guzzi and A.~Tonero, \emph{{Top-quark cross sections and
  distributions at approximate N$^3$LO}},
  \href{https://doi.org/10.1103/PhysRevD.108.054012}{\emph{Phys. Rev. D}
  {\bfseries 108} (2023) 054012}
  [\href{https://arxiv.org/abs/2306.06166}{{\ttfamily 2306.06166}}].

\bibitem{ATLAS:2019hxz}
{\scshape ATLAS} collaboration, \emph{{Measurements of top-quark pair
  differential and double-differential cross-sections in the $\ell$+jets
  channel with $pp$ collisions at $\sqrt{s}=13$ TeV using the ATLAS detector}},
  \href{https://doi.org/10.1140/epjc/s10052-019-7525-6}{\emph{Eur. Phys. J. C}
  {\bfseries 79} (2019) 1028}
  [\href{https://arxiv.org/abs/1908.07305}{{\ttfamily 1908.07305}}].

\bibitem{ATLAS:2020ccu}
{\scshape ATLAS} collaboration, \emph{{Measurements of top-quark pair single-
  and double-differential cross-sections in the all-hadronic channel in $pp$
  collisions at $\sqrt{s}=13~\textrm{TeV}$ using the ATLAS detector}},
  \href{https://doi.org/10.1007/JHEP01(2021)033}{\emph{JHEP} {\bfseries 01}
  (2021) 033} [\href{https://arxiv.org/abs/2006.09274}{{\ttfamily
  2006.09274}}].

\bibitem{CMS:2018adi}
{\scshape CMS} collaboration, \emph{{Measurements of $\mathrm{t\overline{t}}$
  differential cross sections in proton-proton collisions at $\sqrt{s}=$ 13 TeV
  using events containing two leptons}},
  \href{https://doi.org/10.1007/JHEP02(2019)149}{\emph{JHEP} {\bfseries 02}
  (2019) 149} [\href{https://arxiv.org/abs/1811.06625}{{\ttfamily
  1811.06625}}].

\bibitem{CMS:2021vhb}
{\scshape CMS} collaboration, \emph{{Measurement of differential $t \bar t$
  production cross sections in the full kinematic range using lepton+jets
  events from proton-proton collisions at $\sqrt {s}$ = 13\,\,TeV}},
  \href{https://doi.org/10.1103/PhysRevD.104.092013}{\emph{Phys. Rev. D}
  {\bfseries 104} (2021) 092013}
  [\href{https://arxiv.org/abs/2108.02803}{{\ttfamily 2108.02803}}].

\bibitem{Schmidt:2018hvu}
C.~Schmidt, J.~Pumplin, C.~P. Yuan and P.~Yuan, \emph{{Updating and optimizing
  error parton distribution function sets in the Hessian approach}},
  \href{https://doi.org/10.1103/PhysRevD.98.094005}{\emph{Phys. Rev.}
  {\bfseries D98} (2018) 094005}
  [\href{https://arxiv.org/abs/1806.07950}{{\ttfamily 1806.07950}}].

\bibitem{Hou:2019gfw}
T.-J. Hou, Z.~Yu, S.~Dulat, C.~Schmidt and C.~P. Yuan, \emph{{Updating and
  optimizing error parton distribution function sets in the Hessian approach.
  II.}}, \href{https://doi.org/10.1103/PhysRevD.100.114024}{\emph{Phys. Rev. D}
  {\bfseries 100} (2019) 114024}
  [\href{https://arxiv.org/abs/1907.12177}{{\ttfamily 1907.12177}}].

\bibitem{Britzger:2015ksm}
D.~Britzger, G.~S. Klaus~Rabbertz, F.~Stober and M.~Wobisch, \emph{{Recent
  Developments of the fastNLO Toolkit}},
  \href{https://doi.org/10.22323/1.247.0055}{\emph{PoS} {\bfseries DIS2015}
  (2015) 055}.

\bibitem{Czakon:2017dip}
M.~Czakon, D.~Heymes and A.~Mitov, \emph{{fastNLO tables for NNLO top-quark
  pair differential distributions}},
  \href{https://arxiv.org/abs/1704.08551}{{\ttfamily 1704.08551}}.

\bibitem{repo}
``{Repository with $t\bar{t}$ fastNNLO tables}.''
  \url{https://www.precision.hep.phy.cam.ac.uk/results/ttbar-fastnlo/}.

\bibitem{Czakon:2016dgf}
M.~Czakon, D.~Heymes and A.~Mitov, \emph{{Dynamical scales for multi-TeV
  top-pair production at the LHC}},
  \href{https://doi.org/10.1007/JHEP04(2017)071}{\emph{JHEP} {\bfseries 04}
  (2017) 071} [\href{https://arxiv.org/abs/1606.03350}{{\ttfamily
  1606.03350}}].

\bibitem{Grazzini:2017mhc}
M.~Grazzini, S.~Kallweit and M.~Wiesemann, \emph{{Fully differential NNLO
  computations with MATRIX}},
  \href{https://doi.org/10.1140/epjc/s10052-018-5771-7}{\emph{Eur. Phys. J. C}
  {\bfseries 78} (2018) 537}
  [\href{https://arxiv.org/abs/1711.06631}{{\ttfamily 1711.06631}}].

\bibitem{Catani:2019iny}
S.~Catani, S.~Devoto, M.~Grazzini, S.~Kallweit, J.~Mazzitelli and H.~Sargsyan,
  \emph{{Top-quark pair hadroproduction at next-to-next-to-leading order in
  QCD}}, \href{https://doi.org/10.1103/PhysRevD.99.051501}{\emph{Phys. Rev. D}
  {\bfseries 99} (2019) 051501}
  [\href{https://arxiv.org/abs/1901.04005}{{\ttfamily 1901.04005}}].

\bibitem{Matrixrepo}
``{Repository with \texttt{MATRIX} code}.'' \url{https://matrix.hepforge.org/}.

\bibitem{Carli:2010rw}
T.~Carli, D.~Clements, A.~Cooper-Sarkar, C.~Gwenlan, G.~P. Salam, F.~Siegert
  et~al., \emph{{A posteriori inclusion of parton density functions in NLO QCD
  final-state calculations at hadron colliders: The APPLGRID Project}},
  \href{https://doi.org/10.1140/epjc/s10052-010-1255-0}{\emph{Eur. Phys. J. C}
  {\bfseries 66} (2010) 503} [\href{https://arxiv.org/abs/0911.2985}{{\ttfamily
  0911.2985}}].

\bibitem{Campbell:2015qma}
J.~M. Campbell, R.~K. Ellis and W.~T. Giele, \emph{{A Multi-Threaded Version of
  MCFM}}, \href{https://doi.org/10.1140/epjc/s10052-015-3461-2}{\emph{Eur.
  Phys. J. C} {\bfseries 75} (2015) 246}
  [\href{https://arxiv.org/abs/1503.06182}{{\ttfamily 1503.06182}}].

\bibitem{Pagani:2016caq}
D.~Pagani, I.~Tsinikos and M.~Zaro, \emph{{The impact of the photon PDF and
  electroweak corrections on $t \bar{t}$ distributions}},
  \href{https://doi.org/10.1140/epjc/s10052-016-4318-z}{\emph{Eur. Phys. J. C}
  {\bfseries 76} (2016) 479}
  [\href{https://arxiv.org/abs/1606.01915}{{\ttfamily 1606.01915}}].

\bibitem{Campbell:2016dks}
J.~M. Campbell, D.~Wackeroth and J.~Zhou, \emph{{Study of weak corrections to
  Drell-Yan, top-quark pair, and dijet production at high energies with MCFM}},
  \href{https://doi.org/10.1103/PhysRevD.94.093009}{\emph{Phys. Rev. D}
  {\bfseries 94} (2016) 093009}
  [\href{https://arxiv.org/abs/1608.03356}{{\ttfamily 1608.03356}}].

\bibitem{Nason:1987xz}
P.~Nason, S.~Dawson and R.~K. Ellis, \emph{{The Total Cross-Section for the
  Production of Heavy Quarks in Hadronic Collisions}},
  \href{https://doi.org/10.1016/0550-3213(88)90422-1}{\emph{Nucl. Phys. B}
  {\bfseries 303} (1988) 607}.

\bibitem{Beenakker:1988bq}
W.~Beenakker, H.~Kuijf, W.~L. van Neerven and J.~Smith, \emph{{QCD Corrections
  to Heavy Quark Production in p anti-p Collisions}},
  \href{https://doi.org/10.1103/PhysRevD.40.54}{\emph{Phys. Rev. D} {\bfseries
  40} (1989) 54}.

\bibitem{Nason:1989zy}
P.~Nason, S.~Dawson and R.~K. Ellis, \emph{{The One Particle Inclusive
  Differential Cross-Section for Heavy Quark Production in Hadronic
  Collisions}}, \href{https://doi.org/10.1016/0550-3213(89)90286-1}{\emph{Nucl.
  Phys. B} {\bfseries 327} (1989) 49}.

\bibitem{Meng:1989rp}
R.~Meng, G.~A. Schuler, J.~Smith and W.~L. van Neerven, \emph{{Simple Formulae
  for the Order $\alpha_s^3$ {QCD} Corrections to the Reaction $p \bar{p} \to Q
  \bar{Q}$ X}}, \href{https://doi.org/10.1016/0550-3213(90)90352-E}{\emph{Nucl.
  Phys. B} {\bfseries 339} (1990) 325}.

\bibitem{Beenakker:1990maa}
W.~Beenakker, W.~L. van Neerven, R.~Meng, G.~A. Schuler and J.~Smith,
  \emph{{QCD corrections to heavy quark production in hadron hadron
  collisions}},
  \href{https://doi.org/10.1016/S0550-3213(05)80032-X}{\emph{Nucl. Phys. B}
  {\bfseries 351} (1991) 507}.

\bibitem{Mangano:1991jk}
M.~L. Mangano, P.~Nason and G.~Ridolfi, \emph{{Heavy quark correlations in
  hadron collisions at next-to-leading order}},
  \href{https://doi.org/10.1016/0550-3213(92)90435-E}{\emph{Nucl. Phys. B}
  {\bfseries 373} (1992) 295}.

\bibitem{Barnreuther:2012wtj}
P.~B\"arnreuther, M.~Czakon and A.~Mitov, \emph{{Percent Level Precision
  Physics at the Tevatron: First Genuine NNLO QCD Corrections to $q \bar{q} \to
  t \bar{t} + X$}},
  \href{https://doi.org/10.1103/PhysRevLett.109.132001}{\emph{Phys. Rev. Lett.}
  {\bfseries 109} (2012) 132001}
  [\href{https://arxiv.org/abs/1204.5201}{{\ttfamily 1204.5201}}].

\bibitem{Czakon:2012zr}
M.~Czakon and A.~Mitov, \emph{{NNLO corrections to top-pair production at
  hadron colliders: the all-fermionic scattering channels}},
  \href{https://doi.org/10.1007/JHEP12(2012)054}{\emph{JHEP} {\bfseries 12}
  (2012) 054} [\href{https://arxiv.org/abs/1207.0236}{{\ttfamily 1207.0236}}].

\bibitem{Czakon:2012pz}
M.~Czakon and A.~Mitov, \emph{{NNLO corrections to top pair production at
  hadron colliders: the quark-gluon reaction}},
  \href{https://doi.org/10.1007/JHEP01(2013)080}{\emph{JHEP} {\bfseries 01}
  (2013) 080} [\href{https://arxiv.org/abs/1210.6832}{{\ttfamily 1210.6832}}].

\bibitem{Czakon:2013goa}
M.~Czakon, P.~Fiedler and A.~Mitov, \emph{{Total Top-Quark Pair-Production
  Cross Section at Hadron Colliders Through $O(\alpha^4_S)$}},
  \href{https://doi.org/10.1103/PhysRevLett.110.252004}{\emph{Phys. Rev. Lett.}
  {\bfseries 110} (2013) 252004}
  [\href{https://arxiv.org/abs/1303.6254}{{\ttfamily 1303.6254}}].

\bibitem{Czakon:2015owf}
M.~Czakon, D.~Heymes and A.~Mitov, \emph{{High-precision differential
  predictions for top-quark pairs at the LHC}},
  \href{https://doi.org/10.1103/PhysRevLett.116.082003}{\emph{Phys. Rev. Lett.}
  {\bfseries 116} (2016) 082003}
  [\href{https://arxiv.org/abs/1511.00549}{{\ttfamily 1511.00549}}].

\bibitem{Czakon:2016ckf}
M.~Czakon, P.~Fiedler, D.~Heymes and A.~Mitov, \emph{{NNLO QCD predictions for
  fully-differential top-quark pair production at the Tevatron}},
  \href{https://doi.org/10.1007/JHEP05(2016)034}{\emph{JHEP} {\bfseries 05}
  (2016) 034} [\href{https://arxiv.org/abs/1601.05375}{{\ttfamily
  1601.05375}}].

\bibitem{Catani:2019hip}
S.~Catani, S.~Devoto, M.~Grazzini, S.~Kallweit and J.~Mazzitelli,
  \emph{{Top-quark pair production at the LHC: Fully differential QCD
  predictions at NNLO}},
  \href{https://doi.org/10.1007/JHEP07(2019)100}{\emph{JHEP} {\bfseries 07}
  (2019) 100} [\href{https://arxiv.org/abs/1906.06535}{{\ttfamily
  1906.06535}}].

\bibitem{Catani:2020tko}
S.~Catani, S.~Devoto, M.~Grazzini, S.~Kallweit and J.~Mazzitelli,
  \emph{{Top-quark pair hadroproduction at NNLO: differential predictions with
  the $\overline{MS}$ mass}},
  \href{https://doi.org/10.1007/JHEP08(2020)027}{\emph{JHEP} {\bfseries 08}
  (2020) 027} [\href{https://arxiv.org/abs/2005.00557}{{\ttfamily
  2005.00557}}].

\bibitem{Beenakker:1993yr}
W.~Beenakker, A.~Denner, W.~Hollik, R.~Mertig, T.~Sack and D.~Wackeroth,
  \emph{{Electroweak one loop contributions to top pair production in hadron
  colliders}}, \href{https://doi.org/10.1016/0550-3213(94)90454-5}{\emph{Nucl.
  Phys. B} {\bfseries 411} (1994) 343}.

\bibitem{Bernreuther:2005is}
W.~Bernreuther, M.~F\"ucker and Z.~G. Si, \emph{{Mixed QCD and weak corrections
  to top quark pair production at hadron colliders}},
  \href{https://doi.org/10.1016/j.physletb.2005.11.056}{\emph{Phys. Lett. B}
  {\bfseries 633} (2006) 54}
  [\href{https://arxiv.org/abs/hep-ph/0508091}{{\ttfamily hep-ph/0508091}}].

\bibitem{Kuhn:2005it}
J.~H. Kuhn, A.~Scharf and P.~Uwer, \emph{{Electroweak corrections to top-quark
  pair production in quark-antiquark annihilation}},
  \href{https://doi.org/10.1140/epjc/s2005-02423-6}{\emph{Eur. Phys. J. C}
  {\bfseries 45} (2006) 139}
  [\href{https://arxiv.org/abs/hep-ph/0508092}{{\ttfamily hep-ph/0508092}}].

\bibitem{Bernreuther:2006vg}
W.~Bernreuther, M.~Fuecker and Z.-G. Si, \emph{{Weak interaction corrections to
  hadronic top quark pair production}},
  \href{https://doi.org/10.1103/PhysRevD.74.113005}{\emph{Phys. Rev. D}
  {\bfseries 74} (2006) 113005}
  [\href{https://arxiv.org/abs/hep-ph/0610334}{{\ttfamily hep-ph/0610334}}].

\bibitem{Kuhn:2006vh}
J.~H. Kuhn, A.~Scharf and P.~Uwer, \emph{{Electroweak effects in top-quark pair
  production at hadron colliders}},
  \href{https://doi.org/10.1140/epjc/s10052-007-0275-x}{\emph{Eur. Phys. J. C}
  {\bfseries 51} (2007) 37}
  [\href{https://arxiv.org/abs/hep-ph/0610335}{{\ttfamily hep-ph/0610335}}].

\bibitem{Hollik:2007sw}
W.~Hollik and M.~Kollar, \emph{{NLO QED contributions to top-pair production at
  hadron collider}},
  \href{https://doi.org/10.1103/PhysRevD.77.014008}{\emph{Phys. Rev. D}
  {\bfseries 77} (2008) 014008}
  [\href{https://arxiv.org/abs/0708.1697}{{\ttfamily 0708.1697}}].

\bibitem{Bernreuther:2008md}
W.~Bernreuther, M.~Fucker and Z.-G. Si, \emph{{Weak interaction corrections to
  hadronic top quark pair production: Contributions from quark-gluon and b
  anti-b induced reactions}},
  \href{https://doi.org/10.1103/PhysRevD.78.017503}{\emph{Phys. Rev. D}
  {\bfseries 78} (2008) 017503}
  [\href{https://arxiv.org/abs/0804.1237}{{\ttfamily 0804.1237}}].

\bibitem{Bernreuther:2010ny}
W.~Bernreuther and Z.-G. Si, \emph{{Distributions and correlations for top
  quark pair production and decay at the Tevatron and LHC.}},
  \href{https://doi.org/10.1016/j.nuclphysb.2010.05.001}{\emph{Nucl. Phys. B}
  {\bfseries 837} (2010) 90} [\href{https://arxiv.org/abs/1003.3926}{{\ttfamily
  1003.3926}}].

\bibitem{Hollik:2011ps}
W.~Hollik and D.~Pagani, \emph{{The electroweak contribution to the top quark
  forward-backward asymmetry at the Tevatron}},
  \href{https://doi.org/10.1103/PhysRevD.84.093003}{\emph{Phys. Rev. D}
  {\bfseries 84} (2011) 093003}
  [\href{https://arxiv.org/abs/1107.2606}{{\ttfamily 1107.2606}}].

\bibitem{Kuhn:2011ri}
J.~H. Kuhn and G.~Rodrigo, \emph{{Charge asymmetries of top quarks at hadron
  colliders revisited}},
  \href{https://doi.org/10.1007/JHEP01(2012)063}{\emph{JHEP} {\bfseries 01}
  (2012) 063} [\href{https://arxiv.org/abs/1109.6830}{{\ttfamily 1109.6830}}].

\bibitem{Manohar:2012rs}
A.~V. Manohar and M.~Trott, \emph{{Electroweak Sudakov Corrections and the Top
  Quark Forward-Backward Asymmetry}},
  \href{https://doi.org/10.1016/j.physletb.2012.04.013}{\emph{Phys. Lett. B}
  {\bfseries 711} (2012) 313}
  [\href{https://arxiv.org/abs/1201.3926}{{\ttfamily 1201.3926}}].

\bibitem{Bernreuther:2012sx}
W.~Bernreuther and Z.-G. Si, \emph{{Top quark and leptonic charge asymmetries
  for the Tevatron and LHC}},
  \href{https://doi.org/10.1103/PhysRevD.86.034026}{\emph{Phys. Rev. D}
  {\bfseries 86} (2012) 034026}
  [\href{https://arxiv.org/abs/1205.6580}{{\ttfamily 1205.6580}}].

\bibitem{Kuhn:2013zoa}
J.~H. K\"uhn, A.~Scharf and P.~Uwer, \emph{{Weak Interactions in Top-Quark Pair
  Production at Hadron Colliders: An Update}},
  \href{https://doi.org/10.1103/PhysRevD.91.014020}{\emph{Phys. Rev. D}
  {\bfseries 91} (2015) 014020}
  [\href{https://arxiv.org/abs/1305.5773}{{\ttfamily 1305.5773}}].

\bibitem{Campbell:2015vua}
J.~M. Campbell, D.~Wackeroth and J.~Zhou, \emph{{Electroweak Corrections at the
  LHC with MCFM}}, \href{https://doi.org/10.22323/1.247.0130}{\emph{PoS}
  {\bfseries DIS2015} (2015) 130}
  [\href{https://arxiv.org/abs/1508.06247}{{\ttfamily 1508.06247}}].

\bibitem{Denner:2016jyo}
A.~Denner and M.~Pellen, \emph{{NLO electroweak corrections to off-shell
  top-antitop production with leptonic decays at the LHC}},
  \href{https://doi.org/10.1007/JHEP08(2016)155}{\emph{JHEP} {\bfseries 08}
  (2016) 155} [\href{https://arxiv.org/abs/1607.05571}{{\ttfamily
  1607.05571}}].

\bibitem{Czakon:2017wor}
M.~Czakon, D.~Heymes, A.~Mitov, D.~Pagani, I.~Tsinikos and M.~Zaro,
  \emph{{Top-pair production at the LHC through NNLO QCD and NLO EW}},
  \href{https://doi.org/10.1007/JHEP10(2017)186}{\emph{JHEP} {\bfseries 10}
  (2017) 186} [\href{https://arxiv.org/abs/1705.04105}{{\ttfamily
  1705.04105}}].

\bibitem{Carrazza:2020gss}
S.~Carrazza, E.~R. Nocera, C.~Schwan and M.~Zaro, \emph{{PineAPPL: combining EW
  and QCD corrections for fast evaluation of LHC processes}},
  \href{https://doi.org/10.1007/JHEP12(2020)108}{\emph{JHEP} {\bfseries 12}
  (2020) 108} [\href{https://arxiv.org/abs/2008.12789}{{\ttfamily
  2008.12789}}].

\bibitem{Sterman:1986aj}
G.~F. Sterman, \emph{{Summation of Large Corrections to Short Distance Hadronic
  Cross-Sections}},
  \href{https://doi.org/10.1016/0550-3213(87)90258-6}{\emph{Nucl. Phys. B}
  {\bfseries 281} (1987) 310}.

\bibitem{Catani:1989ne}
S.~Catani and L.~Trentadue, \emph{{Resummation of the QCD Perturbative Series
  for Hard Processes}},
  \href{https://doi.org/10.1016/0550-3213(89)90273-3}{\emph{Nucl. Phys. B}
  {\bfseries 327} (1989) 323}.

\bibitem{Kidonakis:1996aq}
N.~Kidonakis and G.~F. Sterman, \emph{{Subleading logarithms in QCD hard
  scattering}}, \href{https://doi.org/10.1016/0370-2693(96)01080-5}{\emph{Phys.
  Lett. B} {\bfseries 387} (1996) 867}.

\bibitem{Kidonakis:1997gm}
N.~Kidonakis and G.~F. Sterman, \emph{{Resummation for QCD hard scattering}},
  \href{https://doi.org/10.1016/S0550-3213(97)00506-3}{\emph{Nucl. Phys. B}
  {\bfseries 505} (1997) 321}
  [\href{https://arxiv.org/abs/hep-ph/9705234}{{\ttfamily hep-ph/9705234}}].

\bibitem{Bonciani:1998vc}
R.~Bonciani, S.~Catani, M.~L. Mangano and P.~Nason, \emph{{NLL resummation of
  the heavy quark hadroproduction cross-section}},
  \href{https://doi.org/10.1016/S0550-3213(98)00335-6}{\emph{Nucl. Phys. B}
  {\bfseries 529} (1998) 424}
  [\href{https://arxiv.org/abs/hep-ph/9801375}{{\ttfamily hep-ph/9801375}}].

\bibitem{Laenen:1998qw}
E.~Laenen, G.~Oderda and G.~F. Sterman, \emph{{Resummation of threshold
  corrections for single particle inclusive cross-sections}},
  \href{https://doi.org/10.1016/S0370-2693(98)00960-5}{\emph{Phys. Lett. B}
  {\bfseries 438} (1998) 173}
  [\href{https://arxiv.org/abs/hep-ph/9806467}{{\ttfamily hep-ph/9806467}}].

\bibitem{Kidonakis:2000ui}
N.~Kidonakis, \emph{{High order corrections and subleading logarithms for top
  quark production}},
  \href{https://doi.org/10.1103/PhysRevD.64.014009}{\emph{Phys. Rev. D}
  {\bfseries 64} (2001) 014009}
  [\href{https://arxiv.org/abs/hep-ph/0010002}{{\ttfamily hep-ph/0010002}}].

\bibitem{Bauer:2000yr}
C.~W. Bauer, S.~Fleming, D.~Pirjol and I.~W. Stewart, \emph{{An Effective field
  theory for collinear and soft gluons: Heavy to light decays}},
  \href{https://doi.org/10.1103/PhysRevD.63.114020}{\emph{Phys. Rev. D}
  {\bfseries 63} (2001) 114020}
  [\href{https://arxiv.org/abs/hep-ph/0011336}{{\ttfamily hep-ph/0011336}}].

\bibitem{Kidonakis:2001nj}
N.~Kidonakis, E.~Laenen, S.~Moch and R.~Vogt, \emph{{Sudakov resummation and
  finite order expansions of heavy quark hadroproduction cross-sections}},
  \href{https://doi.org/10.1103/PhysRevD.64.114001}{\emph{Phys. Rev. D}
  {\bfseries 64} (2001) 114001}
  [\href{https://arxiv.org/abs/hep-ph/0105041}{{\ttfamily hep-ph/0105041}}].

\bibitem{Bauer:2001yt}
C.~W. Bauer, D.~Pirjol and I.~W. Stewart, \emph{{Soft collinear factorization
  in effective field theory}},
  \href{https://doi.org/10.1103/PhysRevD.65.054022}{\emph{Phys. Rev. D}
  {\bfseries 65} (2002) 054022}
  [\href{https://arxiv.org/abs/hep-ph/0109045}{{\ttfamily hep-ph/0109045}}].

\bibitem{Beneke:2002ph}
M.~Beneke, A.~P. Chapovsky, M.~Diehl and T.~Feldmann, \emph{{Soft collinear
  effective theory and heavy to light currents beyond leading power}},
  \href{https://doi.org/10.1016/S0550-3213(02)00687-9}{\emph{Nucl. Phys. B}
  {\bfseries 643} (2002) 431}
  [\href{https://arxiv.org/abs/hep-ph/0206152}{{\ttfamily hep-ph/0206152}}].

\bibitem{Kidonakis:2003qe}
N.~Kidonakis and R.~Vogt, \emph{{Next-to-next-to-leading order soft gluon
  corrections in top quark hadroproduction}},
  \href{https://doi.org/10.1103/PhysRevD.68.114014}{\emph{Phys. Rev. D}
  {\bfseries 68} (2003) 114014}
  [\href{https://arxiv.org/abs/hep-ph/0308222}{{\ttfamily hep-ph/0308222}}].

\bibitem{Kidonakis:2008mu}
N.~Kidonakis and R.~Vogt, \emph{{The Theoretical top quark cross section at the
  Tevatron and the LHC}},
  \href{https://doi.org/10.1103/PhysRevD.78.074005}{\emph{Phys. Rev. D}
  {\bfseries 78} (2008) 074005}
  [\href{https://arxiv.org/abs/0805.3844}{{\ttfamily 0805.3844}}].

\bibitem{Kidonakis:2009ev}
N.~Kidonakis, \emph{{Two-loop soft anomalous dimensions and NNLL resummation
  for heavy quark production}},
  \href{https://doi.org/10.1103/PhysRevLett.102.232003}{\emph{Phys. Rev. Lett.}
  {\bfseries 102} (2009) 232003}
  [\href{https://arxiv.org/abs/0903.2561}{{\ttfamily 0903.2561}}].

\bibitem{Czakon:2009zw}
M.~Czakon, A.~Mitov and G.~F. Sterman, \emph{{Threshold Resummation for
  Top-Pair Hadroproduction to Next-to-Next-to-Leading Log}},
  \href{https://doi.org/10.1103/PhysRevD.80.074017}{\emph{Phys. Rev. D}
  {\bfseries 80} (2009) 074017}
  [\href{https://arxiv.org/abs/0907.1790}{{\ttfamily 0907.1790}}].

\bibitem{Ahrens:2009uz}
V.~Ahrens, A.~Ferroglia, M.~Neubert, B.~D. Pecjak and L.~L. Yang,
  \emph{{Threshold expansion at order $\alpha_s^4$ for the $t \bar t$ invariant
  mass distribution at hadron colliders}},
  \href{https://doi.org/10.1016/j.physletb.2010.03.048}{\emph{Phys. Lett. B}
  {\bfseries 687} (2010) 331}
  [\href{https://arxiv.org/abs/0912.3375}{{\ttfamily 0912.3375}}].

\bibitem{Ahrens:2010zv}
V.~Ahrens, A.~Ferroglia, M.~Neubert, B.~D. Pecjak and L.~L. Yang,
  \emph{{Renormalization-Group Improved Predictions for Top-Quark Pair
  Production at Hadron Colliders}},
  \href{https://doi.org/10.1007/JHEP09(2010)097}{\emph{JHEP} {\bfseries 09}
  (2010) 097} [\href{https://arxiv.org/abs/1003.5827}{{\ttfamily 1003.5827}}].

\bibitem{Aliev:2010zk}
M.~Aliev, H.~Lacker, U.~Langenfeld, S.~Moch, P.~Uwer and M.~Wiedermann,
  \emph{{HATHOR: HAdronic Top and Heavy quarks crOss section calculatoR}},
  \href{https://doi.org/10.1016/j.cpc.2010.12.040}{\emph{Comput. Phys. Commun.}
  {\bfseries 182} (2011) 1034}
  [\href{https://arxiv.org/abs/1007.1327}{{\ttfamily 1007.1327}}].

\bibitem{Kidonakis:2010dk}
N.~Kidonakis, \emph{{Next-to-next-to-leading soft-gluon corrections for the top
  quark cross section and transverse momentum distribution}},
  \href{https://doi.org/10.1103/PhysRevD.82.114030}{\emph{Phys. Rev. D}
  {\bfseries 82} (2010) 114030}
  [\href{https://arxiv.org/abs/1009.4935}{{\ttfamily 1009.4935}}].

\bibitem{Ahrens:2011mw}
V.~Ahrens, A.~Ferroglia, M.~Neubert, B.~D. Pecjak and L.-L. Yang,
  \emph{{RG-improved single-particle inclusive cross sections and
  forward-backward asymmetry in $t\bar t$ production at hadron colliders}},
  \href{https://doi.org/10.1007/JHEP09(2011)070}{\emph{JHEP} {\bfseries 09}
  (2011) 070} [\href{https://arxiv.org/abs/1103.0550}{{\ttfamily 1103.0550}}].

\bibitem{Kidonakis:2011zn}
N.~Kidonakis, \emph{{The top quark rapidity distribution and forward-backward
  asymmetry}}, \href{https://doi.org/10.1103/PhysRevD.84.011504}{\emph{Phys.
  Rev. D} {\bfseries 84} (2011) 011504}
  [\href{https://arxiv.org/abs/1105.5167}{{\ttfamily 1105.5167}}].

\bibitem{Ahrens:2011px}
V.~Ahrens, A.~Ferroglia, M.~Neubert, B.~D. Pecjak and L.~L. Yang,
  \emph{{Precision predictions for the $t \bar t$ production cross section at
  hadron colliders}},
  \href{https://doi.org/10.1016/j.physletb.2011.07.058}{\emph{Phys. Lett. B}
  {\bfseries 703} (2011) 135}
  [\href{https://arxiv.org/abs/1105.5824}{{\ttfamily 1105.5824}}].

\bibitem{Beneke:2011mq}
M.~Beneke, P.~Falgari, S.~Klein and C.~Schwinn, \emph{{Hadronic top-quark pair
  production with NNLL threshold resummation}},
  \href{https://doi.org/10.1016/j.nuclphysb.2011.10.021}{\emph{Nucl. Phys. B}
  {\bfseries 855} (2012) 695}
  [\href{https://arxiv.org/abs/1109.1536}{{\ttfamily 1109.1536}}].

\bibitem{Cacciari:2011hy}
M.~Cacciari, M.~Czakon, M.~Mangano, A.~Mitov and P.~Nason, \emph{{Top-pair
  production at hadron colliders with next-to-next-to-leading logarithmic
  soft-gluon resummation}},
  \href{https://doi.org/10.1016/j.physletb.2012.03.013}{\emph{Phys. Lett. B}
  {\bfseries 710} (2012) 612}
  [\href{https://arxiv.org/abs/1111.5869}{{\ttfamily 1111.5869}}].

\bibitem{Czakon:2011xx}
M.~Czakon and A.~Mitov, \emph{{Top++: A Program for the Calculation of the
  Top-Pair Cross-Section at Hadron Colliders}},
  \href{https://doi.org/10.1016/j.cpc.2014.06.021}{\emph{Comput. Phys. Commun.}
  {\bfseries 185} (2014) 2930}
  [\href{https://arxiv.org/abs/1112.5675}{{\ttfamily 1112.5675}}].

\bibitem{Ferroglia:2012ku}
A.~Ferroglia, B.~D. Pecjak and L.~L. Yang, \emph{{Soft-gluon resummation for
  boosted top-quark production at hadron colliders}},
  \href{https://doi.org/10.1103/PhysRevD.86.034010}{\emph{Phys. Rev. D}
  {\bfseries 86} (2012) 034010}
  [\href{https://arxiv.org/abs/1205.3662}{{\ttfamily 1205.3662}}].

\bibitem{Beneke:2012wb}
M.~Beneke, P.~Falgari, S.~Klein, J.~Piclum, C.~Schwinn, M.~Ubiali et~al.,
  \emph{{Inclusive Top-Pair Production Phenomenology with TOPIXS}},
  \href{https://doi.org/10.1007/JHEP07(2012)194}{\emph{JHEP} {\bfseries 07}
  (2012) 194} [\href{https://arxiv.org/abs/1206.2454}{{\ttfamily 1206.2454}}].

\bibitem{Ferroglia:2012uy}
A.~Ferroglia, B.~D. Pecjak, L.~L. Yang, B.~D. Pecjak and L.~L. Yang, \emph{{The
  NNLO soft function for the pair invariant mass distribution of boosted top
  quarks}}, \href{https://doi.org/10.1007/JHEP10(2012)180}{\emph{JHEP}
  {\bfseries 10} (2012) 180} [\href{https://arxiv.org/abs/1207.4798}{{\ttfamily
  1207.4798}}].

\bibitem{Kidonakis:2012rm}
N.~Kidonakis, \emph{{NNLL threshold resummation for top-pair and single-top
  production}}, \href{https://doi.org/10.1134/S1063779614040091}{\emph{Phys.
  Part. Nucl.} {\bfseries 45} (2014) 714}
  [\href{https://arxiv.org/abs/1210.7813}{{\ttfamily 1210.7813}}].

\bibitem{Ferroglia:2013awa}
A.~Ferroglia, S.~Marzani, B.~D. Pecjak and L.~L. Yang, \emph{{Boosted top
  production: factorization and resummation for single-particle inclusive
  distributions}}, \href{https://doi.org/10.1007/JHEP01(2014)028}{\emph{JHEP}
  {\bfseries 01} (2014) 028} [\href{https://arxiv.org/abs/1310.3836}{{\ttfamily
  1310.3836}}].

\bibitem{Kidonakis:2014isa}
N.~Kidonakis, \emph{{NNNLO soft-gluon corrections for the top-antitop pair
  production cross section}},
  \href{https://doi.org/10.1103/PhysRevD.90.014006}{\emph{Phys. Rev. D}
  {\bfseries 90} (2014) 014006}
  [\href{https://arxiv.org/abs/1405.7046}{{\ttfamily 1405.7046}}].

\bibitem{Kidonakis:2014pja}
N.~Kidonakis, \emph{{NNNLO soft-gluon corrections for the top-quark $p_T$ and
  rapidity distributions}},
  \href{https://doi.org/10.1103/PhysRevD.91.031501}{\emph{Phys. Rev. D}
  {\bfseries 91} (2015) 031501}
  [\href{https://arxiv.org/abs/1411.2633}{{\ttfamily 1411.2633}}].

\bibitem{Kidonakis:2015ona}
N.~Kidonakis, \emph{{The top quark forward-backward asymmetry at approximate
  N$^3$LO}}, \href{https://doi.org/10.1103/PhysRevD.91.071502}{\emph{Phys. Rev.
  D} {\bfseries 91} (2015) 071502}
  [\href{https://arxiv.org/abs/1501.01581}{{\ttfamily 1501.01581}}].

\bibitem{Kidonakis:2019yji}
N.~Kidonakis, \emph{{Top-quark double-differential distributions at approximate
  N$^3$LO}}, \href{https://doi.org/10.1103/PhysRevD.101.074006}{\emph{Phys.
  Rev. D} {\bfseries 101} (2020) 074006}
  [\href{https://arxiv.org/abs/1912.10362}{{\ttfamily 1912.10362}}].

\bibitem{Hou:2019efy}
T.-J. Hou et~al., \emph{{New CTEQ global analysis of quantum chromodynamics
  with high-precision data from the LHC}},
  \href{https://doi.org/10.1103/PhysRevD.103.014013}{\emph{Phys. Rev. D}
  {\bfseries 103} (2021) 014013}
  [\href{https://arxiv.org/abs/1912.10053}{{\ttfamily 1912.10053}}].

\bibitem{Bailey:2020ooq}
S.~Bailey, T.~Cridge, L.~A. Harland-Lang, A.~D. Martin and R.~S. Thorne,
  \emph{{Parton distributions from LHC, HERA, Tevatron and fixed target data:
  MSHT20 PDFs}},
  \href{https://doi.org/10.1140/epjc/s10052-021-09057-0}{\emph{Eur. Phys. J. C}
  {\bfseries 81} (2021) 341}
  [\href{https://arxiv.org/abs/2012.04684}{{\ttfamily 2012.04684}}].

\bibitem{McGowan:2022nag}
J.~McGowan, T.~Cridge, L.~A. Harland-Lang and R.~S. Thorne, \emph{{Approximate
  N$^{3}$LO parton distribution functions with theoretical uncertainties:
  MSHT20aN$^3$LO PDFs}},
  \href{https://doi.org/10.1140/epjc/s10052-023-11236-0}{\emph{Eur. Phys. J. C}
  {\bfseries 83} (2023) 185}
  [\href{https://arxiv.org/abs/2207.04739}{{\ttfamily 2207.04739}}].

\bibitem{NNPDF:2021njg}
{\scshape NNPDF} collaboration, \emph{{The path to proton structure at 1\%
  accuracy}}, \href{https://doi.org/10.1140/epjc/s10052-022-10328-7}{\emph{Eur.
  Phys. J. C} {\bfseries 82} (2022) 428}
  [\href{https://arxiv.org/abs/2109.02653}{{\ttfamily 2109.02653}}].

\bibitem{ATLAS:2022jbj}
{\scshape ATLAS} collaboration, \emph{{Measurement of the $ t\overline{t} $
  production cross-section in pp collisions at $ \sqrt{s} $ = 5.02 TeV with the
  ATLAS detector}}, \href{https://doi.org/10.1007/JHEP06(2023)138}{\emph{JHEP}
  {\bfseries 06} (2023) 138}
  [\href{https://arxiv.org/abs/2207.01354}{{\ttfamily 2207.01354}}].

\bibitem{CMS:2021gwv}
{\scshape CMS} collaboration, \emph{{Measurement of the inclusive $
  \mathrm{t}\overline{\mathrm{t}} $ production cross section in proton-proton
  collisions at $ \sqrt{s} $ = 5.02 TeV}},
  \href{https://doi.org/10.1007/JHEP04(2022)144}{\emph{JHEP} {\bfseries 04}
  (2022) 144} [\href{https://arxiv.org/abs/2112.09114}{{\ttfamily
  2112.09114}}].

\bibitem{ATLAS:2022aof}
{\scshape ATLAS, CMS} collaboration, \emph{{Combination of inclusive top-quark
  pair production cross-section measurements using ATLAS and CMS data at $
  \sqrt{s} $ = 7 and 8 TeV}},
  \href{https://doi.org/10.1007/JHEP07(2023)213}{\emph{JHEP} {\bfseries 07}
  (2023) 213} [\href{https://arxiv.org/abs/2205.13830}{{\ttfamily
  2205.13830}}].

\bibitem{ATLAS:2023gsl}
{\scshape ATLAS} collaboration, \emph{{Inclusive and differential
  cross-sections for dilepton $ t\overline{t} $ production measured in $
  \sqrt{s} $ = 13 TeV pp collisions with the ATLAS detector}},
  \href{https://doi.org/10.1007/JHEP07(2023)141}{\emph{JHEP} {\bfseries 07}
  (2023) 141} [\href{https://arxiv.org/abs/2303.15340}{{\ttfamily
  2303.15340}}].

\bibitem{Alwall:2014hca}
J.~Alwall, R.~Frederix, S.~Frixione, V.~Hirschi, F.~Maltoni, O.~Mattelaer
  et~al., \emph{{The automated computation of tree-level and next-to-leading
  order differential cross sections, and their matching to parton shower
  simulations}}, \href{https://doi.org/10.1007/JHEP07(2014)079}{\emph{JHEP}
  {\bfseries 07} (2014) 079} [\href{https://arxiv.org/abs/1405.0301}{{\ttfamily
  1405.0301}}].

\bibitem{Frederix:2018nkq}
R.~Frederix, S.~Frixione, V.~Hirschi, D.~Pagani, H.~S. Shao and M.~Zaro,
  \emph{{The automation of next-to-leading order electroweak calculations}},
  \href{https://doi.org/10.1007/JHEP07(2018)185}{\emph{JHEP} {\bfseries 07}
  (2018) 185} [\href{https://arxiv.org/abs/1804.10017}{{\ttfamily
  1804.10017}}].

\end{thebibliography}

\providecommand{\href}[2]{#2}\begingroup\raggedright\endgroup

\end{document}